\documentclass[12pt,letterpaper]{article}

\usepackage{amssymb,amsfonts,amsmath}
\usepackage{mathtools}
\usepackage{setspace}
\usepackage{sourceserifpro}
\usepackage[T1]{fontenc}
\usepackage{graphicx}
\usepackage[left=1in,right=1in,top=1in,bottom=1in]{geometry}
\usepackage[authoryear]{natbib}

\usepackage{setspace}
\usepackage[authoryear]{natbib}
\usepackage{enumerate}
\usepackage{hyperref}
\usepackage{booktabs}
\usepackage{pdflscape}
\usepackage{tabularx}
\usepackage{bm}
\usepackage{mathtools}
\usepackage{tikz}
\usepackage{subcaption}
\usepackage[flushleft]{threeparttable}
\usepackage{mathrsfs}
\usepackage[shortlabels]{enumitem}

\hypersetup{
	colorlinks=true,
	linkcolor=blue,
	filecolor=magenta,
	urlcolor=cyan,
	citecolor=blue,
}
\newtheorem{theorem}{Theorem}[section]
\newtheorem{lemma}[theorem]{Lemma}

\newtheorem{proposition}[theorem]{Proposition}

\newtheorem{assumption}[theorem]{Assumption}

\newenvironment{proof}[1][Proof]{\begin{trivlist}
		\item[\hskip \labelsep {\bfseries #1}]}{\end{trivlist}}
\newtheorem{definition}[theorem]{Definition}

\newcommand{\ex}{{\bf\sf E}}
\newcommand{\pr}{{\bf\sf P}}
\newcommand{\sfS}{{\bf\sf S}}
\newcommand{\sfT}{{\bf\sf T}}
\newcommand{\cA}{\mathbb{A}} 
\newcommand{\cB}{\mathbb{B}} 
\newcommand{\cX}{\mathbb{X}} 
\newcommand{\cY}{\mathbb{Y}} 
\newcommand{\cZ}{\mathbb{Z}} 

\begin{document}

	\begin{titlepage}
		\title{Equilibrium Defaultable Corporate Debt and 
		Investment\footnote{We thank Lorenzo Garlappi, Bob Goldstein, Zhiguo 
		He, David D.\ Yao and Xunyu Zhou for helpful comments. We alone are 
		responsible for any errors.}}
		
		\author{Hong Chen\thanks{%
				Shanghai Advanced Institute in Finance, Shanghai Jiao Tong
				University, Shanghai, China. E-mail: hchen@saif.sjtu.edu.cn}
			\and Murray Z. Frank\thanks{Department of Finance, University
				of Minnesota, Minneapolis, MN 55455 and SAIF. Email:
				murra280@umn.edu} }

		\date{February 11, 2022}
		
		\maketitle
		
		\begin{abstract}
			\begin{singlespace}
				\noindent
				In dynamic capital structure models with an investor break-even 
				condition, the firm's Bellman equation may not generate a 
				contraction mapping, so the standard existence and uniqueness 
				conditions do not apply. First, we provide an example showing 
				the problem in a classical trade-off model. The firm can issue 
				one-period defaultable debt, invest in capital and pay a 
				dividend. If the firm cannot meet the required debt payment, it 
				is liquidated. Second, we show how to use a dual to the 
				original problem and a change of measure, such that existence 
				and uniqueness can be proved.  In the unique Markov-perfect 
				equilibrium, firm decisions reflect state-dependent capital and 
				debt targets. Our approach may be useful for other dynamic firm 
				models that have an investor break-even condition.
			\end{singlespace}
		\end{abstract}
		
		\thispagestyle{empty} \DeclareGraphicsExtensions{.pdf,.png,.gif,.jpg}
		
	\end{titlepage}

	\clearpage
	\setstretch{1.5}
	
	\setcounter{page}{1}
	
\section{Introduction}

We consider a discrete-time model of a firm that can issue one-period 
defaultable debt, invest in capital and pay a dividend. The firm has a 
concave production function of the capital and production is subject to an 
exogenous random shock. In each period, the firm starts with a capital level, 
a bond payment that is due on old debt. The exogenous state is revealed so 
production revenue becomes known. The firm decides the amount of the new debt 
to be issued or the savings, investment in capital, and dividends to pay. The 
firm must make the promised payment on old debt if it can. If the production 
revenue and new debt are not sufficient to make the promised payment to old
debt, the firm is in default. In default the firm is liquidated and the 
proceeds go to the debt holder. Due to possible default the debt is risky. The
price of debt is determined by the equilibrium requirement that the investor 
must expect to break-even relative to the risk-free interest rate.

The firm maximizes its expected total discounted dividend over an infinite 
horizon or to the time epoch when it enters default (whichever comes first). 
The Bellman equation for the firm problem does not generate a contraction 
mapping. Accordingly, the usual existence and uniqueness proof as in 
\citet{stokey1989recursive} and 
\citet{Puterman1994} does not apply. This is due to the connection between 
default risk and debt pricing. Specifically, the current debt price depends 
on whether the firm will default in the next period or the value function of 
the next period. We provide an alternative approach by introducing a dual 
problem. With a change of measure, the dual problem is shown to be a 
contraction mapping and there exists a unique solution. This shows that the 
Markov-perfect equilibrium (Definition~\ref{Equilibrium-definition}) exists 
and is unique for the original problem. Under the equilibrium, we show that 
the price of debt decreases with the amount of debt issued, and the firm 
never saves. There exists an optimal capital target and bond target. These
depend on the exogenous state. When capital is below target and the cash 
flow is sufficient, the firm issues debt to reach the target.

Our model is a basic capital structure trade-off theory model. It is widely 
believed that trade-off theory provides a good account of the debt and 
investment decisions of large public firms, see \cite{ai2021trade}. That 
belief is largely grounded in a sequence of influential papers including, 
\cite{cooley2001financial, hennessy2007costly, nikolov2021sources}. These 
papers show that numerical simulations of trade-off models with investment 
and financing frictions can generate empirically appealing solutions when 
the parameters are reasonably calibrated or estimated. However, the equilibrium 
theory underpinning that empirical success is less well established than 
generally understood. Without establishing the existence and uniqueness for 
these models, any numerical simulations may not be a good reflection of the 
economic forces at work. If the Markov-perfect equilibrium is not unique, then 
interplay among the equilibria may be important for an understanding of the 
driving forces. If however, the equilibrium exist and is unique, then more 
traditional approaches may suffice \citep{strebulaev2012dynamic}.

Why has the difficulty that we address not already been dealt with in the 
literature? Previous studies have generally sidestepped the issue by making 
specialized assumptions. For example \cite{gomes2001financing} does not require 
that debt investors expect to break-even. \cite{hennessy2007costly} impose a 
debt break-even condition on the debt market, but they have a reduced form 
equity market without a break-even requirement. Conditional on a given interest 
rate, they provide conditions such that the firm's value function exists and 
it satisfies a contraction mapping. However, they also caution that there are 
technical issues such that in both \cite{cooley2001financial} and
\cite{hennessy2007costly} the firm's problem is not necessarily concave, and
their overall equilibrium is not necessarily unique. \cite{nikolov2021sources} 
impose an investor break-even condition, and find that the model provides a 
good account of the decisions of large public firms in Compustat. But they do 
not prove equilibrium existence or uniqueness. Our model does not have 
specialized assumptions to sidestep the technical difficulties. Instead we 
stick with classical assumptions and show how to deal with the associated 
challenges. 

The conceptual difficulty is not a mathematical or a technical quirk. It is due
to the fundamental equilibrium requirement that investors will not knowingly 
invest at an avoidable expected loss. Together with the possibility of firm 
default, this means that the original model does not define a contraction 
mapping, and so the standard justification for existence and uniqueness in
\cite{stokey1989recursive} does not apply to the model. Corresponding issues 
may arise in other dynamic equilibrium settings in which future payoffs depend 
both on the realization of a stochastic process and on decisions of other 
people.

There is a large volume of the literature concerning default, 
particularly, following the last financial crisis; see, for example, 
\cite{glasserman2016,chen2021}, and the references in their works. 
These works are concerned with the contagion effects in a network setting. They 
do not model the lending decision of the debt holder. 

Closely related to our concern about establishing uniqueness in a basic risky
corporate debt model, is found in the literature on sovereign debt
\citep{auclert2016unique, aguiar2019contraction}. Due to the investor
break-even condition, the natural government debt problem does not define a
contraction mapping. Our focus on the dual problem instead of the primal 
problem follows \cite{aguiar2019contraction}. The approach will of course not 
apply if there are multiple equilibria. Multiple equilibria have been found 
in problems with multi-period debt 
\citep{aguiar2019take,dangl2021debt} and in some models with 
financial frictions \citep{hugonnier2015credit}.

In this paper, we show that the model does actually have a unique Markov perfect
equilibrium. To prove this, we begin as in \cite{aguiar2019contraction} by 
obtain a dual to the original problem. But then as in \cite{lippman1975dynamic} 
and \cite{nunen1978} we change the measure. Under the change of measure, the dual 
to the original model satisfies the \cite{blackwell1965discounted} conditions. 
So the original model does actually have a unique equilibrium, even with firm 
default and the investor break-even requirement. Firm default is a common 
concern, so our approach may be useful beyond the specifics of our model. 
\cite{hotchkiss2008bankruptcy} provides a very good review of the default 
literature.  

We examine two variations on the basic model. First, we replace the capital 
and the production function with an exogenous cash flow in the main model. This 
is quite common \citep{modigliani1958cost,leland1994corporate}, particularly in 
continuous time capital structure models. Second, we complicate the main model 
by including capital adjustment costs, which is also quite common in the 
literature \citep{summers1981taxation}. In both cases we maintain the 
assumption of single period debt. In both cases our approach establishes 
existence and uniqueness of the equilibrium. Such models are commonly solved 
numerically. Having a version of the model with an equilibrium that is known 
to be unique, is useful for checking code while developing more complex 
numerical models \citep{judd1998numerical}.

The rest of the paper is organized as follows. The model, the equilibrium 
definition and the main result are presented in section~\ref{sec:DebtPossible}. 
The main result is proved in section~\ref{sec-proof}. The proof starts with a 
counterexample to show that the usual Bellman equation, namely, the primal 
problem, does not give a contraction mapping. Then, we introduce a 
(corresponding) dual problem with the solution to the dual uniquely related to 
the solution to the primal problem. We show that with a change of measure, the 
Bellman equation associated with the dual problem gives rise to a contraction 
mapping and hence has a unique solution. In 
section~\ref{sec-capital-debt-dividend}, we provide some characterizations of 
the equilibrium. In section~\ref{sec:extend}, we consider two alternative 
models which are quite common in the literature. One model replaces the capital 
and the production with an exogenous cash flow, and the other model adds the 
capital adjustment costs. We show that our approach can be used in both cases 
to show the existence and the uniqueness of Markov-perfect equilibrium. 
Section~\ref{sec-conclusion} concludes with a summary and discussions.

\section{The model and the main result}
\label{sec:DebtPossible}
The base model assumptions are classical: the firm objective function is the
expected present value of dividends, there is a linear flow budget constraint
that imposes cash in equals cash out, there is a production function that
depends on capital and a shock, capital depreciates, there is no capital
adjustment cost, firm revenue is taxable with depreciation tax deductible. The
productivity shock is given by a Markov process. Debt is for a single period.
Interest payments are tax deductible and the firm may be in default in some
states. The investor must expect to break-even on the debt purchased from the
firm. If the firm cannot make a promised debt payment, the firm is liquidated
with the proceeds going to the debt holder.

Consider a firm with a concave production function,
\begin{eqnarray}
F(k,z)=zA k^\alpha,\label{eq-productionfunction}
\end{eqnarray}
that depends on capital $k$, a random shock $z$, and fixed parameters $A>0$ and
$\alpha\in(0,1)$. The exogenous shock follows a bounded discrete-time Markov process $\{z_t,t\ge0\}$ with a state space $\cZ\subseteq \Re_+$, where $\Re_+$ is the set of all non-negative real numbers.
\begin{assumption}
\label{ass-bounded-shock}
The Markov process  $\{z_t,t\ge0\}$ has a bounded state space $\cZ$.	That is,
there exists a positive constant $\bar{z}<\infty$ such that $\pr(0\le
z_t\le\bar{z})=1$. Transitions are homogeneous.
\end{assumption}

Unless necessary, we will usually drop the time subscript $t$ for ease of 
presentation. For any (possibly vector) process $\{x_t,t\ge0\}$, when both the 
subscript $t$ (the current period) and the subscript $t+1$ (the next period) 
appear in the same set of equations, we write $x$ for $x_t$ and $x'$ for $x_{t+1}$.  

At the start of a period the firm inherits bonds $b$, and physical capital $k$.
The value of the shock $z$, is revealed. Provided the firm is not in default,
it chooses bonds, dividends, and investment in physical capital denoted
respectively as $b,d, i$. However, if the shock is too bad, then even with the
best decisions the firm is unable to make the promised payment to debt. The
firm enters a permanent default. Then equity gets zero. Debt gets a liquidation
payoff of $L(k')$. When the default happens, it happens after the start of the 
period; so, the capital level $k'$, instead of $k$ is used, to account for 
depreciation.  It is assumed that $L(\cdot)$ is a non-decreasing function.

The price of debt is determined by market equilibrium. The debt investor must
expect to break even. There is an exogenous risk-free interest rate $\rho$,
where  $0<\rho<1$. The price of debt is $q$. If the firm promises to pay $b'$ in the next period, 
it will get $qb'$ in the current period. 
Investors understand that this promise will not be kept if 
it is not feasible to do so. When choosing the amount of debt to issue, the 
firm recognizes that the price of debt depends on the current shock, the 
capital after investment, and the amount of debt issued, $q=q(z, b', 
k')$. These are determined as part of an equilibrium, as shown below.

There is a corporate tax on firm revenue, $\tau$. Depreciation $\delta
k$, and interest payments $(1-q(z,b',k'))b'$, are both
tax  deductible. Investment in physical capital is $i=k'-(1-\delta )k$.

The firm flow budget is,
\begin{equation}
\underbrace{(1-\tau)F(z,k)}_{\substack{ \text{revenue}  \\
		\text{after-tax}}}+\underbrace{\tau(\delta
		k+(1-q)b')}_{%
	\text{tax deduction revenue}}
+\underbrace{qb'-b}_{\substack{ \text{net debt}  \\
		\text{revenue}}}=\underbrace{d}_{\text{dividends}}+\underbrace{i%
,}_{\text{investment}}  \label{eq:FirmBudget-borrow}
\end{equation}
where $q=q(z,b',k')$. As in \cite{strebulaev2012dynamic}, the interest tax 
deduction is received at the same time that the debt is issued. This is a 
common simplification. Suppose instead that the tax deduction is received 
later. Then the interest tax deduction at time $t$ would become $\tau(1-q(z_{t-1},b_{t},k_t))b_{t}$. That would imply that both $z_{t-1}$ 
and $z_{t}$ are in the current budget, complicating the algebra.

The state of the firm is described by a three-dimensional vector $x=(z,b,k)$.
The first component is the exogenous state $z\in\cZ$. The second component is
the (net) debt $b\in\cB=[\underbar{$b$},\bar{b}]$, where 
$\underbar{$b$}\le0\le\bar{b}<\infty$. A negative value of $b$ means savings. 
The finite upper bound $\bar{b}$ imposed on the debt is a No-Ponzi condition. 
The third component is the capital level $k\in\Re_+$. So 
$\cX=\cZ\times\cB\times\Re_+$ is the state space. Let $V(x)\equiv V(z,b,k)$ 
denote the equilibrium value of the firm equity. Note that $V(x)=0$ if the 
firm is in a default state, and $V(x)>0$ if the firm is in a non-default 
state.

The firm's per-period net resources apart from debt, is
$R(z,k)=(1-\tau)F(z,k)+[1-\delta(1-\tau)] k$.
Given the state $(z,b,k)$, the net revenue of the firm is given by
\begin{eqnarray*}
&&	(1-\tau)F(z,k)+\tau\delta k+[\tau+(1-\tau)q(z,b',k')]b^{\prime}-b - i\\
&=&R(z,k)+[\tau+(1-\tau)q(z,b',k')]b^{\prime}-k'-b,
\end{eqnarray*}
Let
\begin{eqnarray}
S(z,k)&=&\sup_{\underbar{${\scriptstyle b}$}\le b'\le \bar{b}
\atop k'\ge(1-\delta)k}\big\{[\tau+(1-\tau)q(z,b',k')]b^{\prime}-k'\big\}.\label{debt-prod-B}
\end{eqnarray}
Then the maximum possible net revenue of the firm is given by
$R(z,k)+S(z,k)-b$. This must be non-negative for the firm to avoid default. 
Notice that if the supremum $\sup$ is not achievable by any $(b',k')$, then 
we would require $R(z,k)+S(z,k)-b$ to be strictly positive to avoid default. 
We show later that this supremum is achievable in the equilibrium. So we use 
the non-strict inequality for simplicity in the definition of the feasible 
set.

We can now define the feasible set of non-default states.
\begin{definition}\label{def-feasible}
	In any equilibrium, the feasible set is defined as,
	\begin{equation*}
	\mathbb{X}_{feas}=\Big\{(z,b,k)\in\cX\ \big| \
		R(z,k) + S(z,k) \ge b%
		\Big\}. 
	\end{equation*}
\end{definition}

Let
\begin{equation*}
\underbar{$b$}(z,k)=\inf\{b\ge\underbar{$b$}:(z,b,k)\in
		\mathbb{X}_{feas}\}\quad\mbox{and}\quad
\bar{b}(z,k)=\sup\{b\le\bar{b}:(z,b,k)\in \mathbb{X}_{feas}\}.
\end{equation*}
Notice that when $k$ is sufficiently large, $\underbar{$b$}(z,k)=\underbar{$b$}$ and
$\bar{b}(z,k)=\bar{b}$. This is due to the fact that $S(z,k)\le \left[\tau+(1-\tau)/(1+\rho)\right]\bar{b}-(1-\delta)k$ (in view of the break-even condition (\ref{eq-upperbound-q})) and then
\begin{eqnarray*}
R(z,k) + S(z,k) \le   (1-\tau)F(z,k)+\tau k +\left[\tau+(1-\tau)/(1+\rho)\right]\bar{b}
\to\infty\mbox{ as }k\to\infty.
\end{eqnarray*}

\smallskip\noindent
\textit{Firm problem.}
Let $x=(z,b,k)\in\cX$ and take $q=q(z,b',k')$ as given.
\begin{eqnarray}
&&\mbox{If $x\in \mathbb{X}_{feas}$, then}\notag\\
\label{eq:FirmBellman-s}
&&\qquad V(x)=\sup_{(b^{\prime },d,i)}\big\{d
+\beta
\mathbf{\mathsf{E}}_z V(x')\big\}, \\
&&\qquad \mbox{subject to}  \notag \\
&& \qquad \qquad (1-\tau)F(z,k)+\tau\delta k+
[\tau+(1-\tau)q(z,b',k')]b^{\prime}
-b =d+i,\notag\\
&& \qquad \qquad k'=(1-\delta)k+i,\notag
\\
&& \qquad \qquad \underbar{$b$}\le b^{\prime }\leq \bar{b},d\geq 0,i\geq 0.\notag\\
&&\mbox{Otherwise, }V(x)=0.\label{eq:FirmBellman-s2}
\end{eqnarray}%
If the firm is viable, the firm is maximizing the expected present value 
of dividends, subject to a flow budget constraint, the investment condition, 
and the bounds on debt, dividends and investment. If the firm is not viable, 
the payoff is zero.

Debt investors will not intentionally invest at a loss. There are many potential 
investors that compete with each other. These considerations are commonly 
represented by the requirement that the investor expects to just break even 
in an equilibrium.

\smallskip\noindent
\textit{Investor break-even condition.} Given the firm value function $V$, the
firm's exogenous state $z$, and the firm's capital level $k'$. If the bond
issue is $b^{\prime} \leq 0$, then $q(z,b',k')=1/(1+\rho)$.
If $b^{\prime } \geq 0$, then
\begin{align}
q(z,b',k')&=\frac{b^{\prime}\mathbf{\mathsf{E}}%
	_{z}1_{\{V(x^{\prime })> 0\}}+L(k^{\prime })\mathbf{\mathsf{E}}%
	_{z}1_{\{V(x^{\prime })=0\}}}{b^{\prime }\left(1+\rho\right)}\qquad
\mbox{ and} \label{eq:BE_basic-s}\\
q(z,b',k')&\le\frac{1}{1+\rho}.\label{eq-upperbound-q}
\end{align}

Condition (\ref{eq-upperbound-q}) means that the debt investor is willing to
pay a price no greater than the price at a risk free rate. After-all, the
investor has access to that risk free rate. Combining the above two conditions,
we have
$[b'-L(k')]\mathbf{\mathsf{E}}_{z}1_{\{V(x^{\prime })=0\}}\ge0$.
This means that at the time of default, the liquidation value of the capital
$L(k')$ must be no more than the debt payment $b'$. We assume that $L(k')$ is
 non-decreasing in $k'$.

Equality (\ref{eq:BE_basic-s}) can be rewritten as
\begin{eqnarray*}
	q(z,b',k')=\frac{b^{\prime}-[b'-L(k^{\prime })]\mathbf{\mathsf{E}}%
		_{z}1_{\{V(x^{\prime })=0\}}}{b^{\prime }\left(1+\rho\right)}.
\end{eqnarray*}
This shows that $q(z,b',k')$ is increasing in the value function $V$.

With the firm problem and the investor break-even conditions in hand, we can now
define the equilibrium.
\begin{definition}
	\label{Equilibrium-definition}
	A Markov-perfect equilibrium is a value function $V$ and a price function
	$q$ such that
	\begin{description}
		\item[(1)] Given $q$,  the value function $V$ satisfies
		the Bellman equation (\ref{eq:FirmBellman-s}) and (\ref{eq:FirmBellman-s2}),
		and the supremum in (\ref{eq:FirmBellman-s}) is attained for some policy;
		\item[(2)] Given $V$, $q$ satisfies the break-even condition
		(\ref{eq:BE_basic-s})-(\ref{eq-upperbound-q}).
	\end{description}
\end{definition}
The Markov Perfect equilibrium is a popular refinement of the Nash equilibrium. 
It is Markov in the sense that all equilibrium decision rules are functions of 
the state variables and do not depend on the rest of the history. The 
equilibrium is Perfect in the sense of being complete and consistent with 
optimizing decisions in all possible future states 
\citep{levhari1980,maskin2001markov}.

\begin{theorem}
	\label{thm-main-theorem}
Under Assumption~\ref{ass-bounded-shock}, there exists a unique Markov-perfect
equilibrium.
\end{theorem}

To justify existence and uniqueness in dynamic corporate capital structure 
models it is common to cite Theorem 9.6 in \citet{stokey1989recursive}. Their 
theorem in turn is justified by \citet{blackwell1965discounted}. In section 
\ref{sub:Primal}, we show by example that due to the investor break-even 
condition, the usual justification does not apply to our model. In section 
\ref{sub:Dual}, we provide an alternative method that does apply.

\section{Proof of the main result}
\label{sec-proof}

We start with an example to show that the obvious operator associated with 
the original firm problem is not a contraction mapping 
(subsection~\ref{sub:Primal}). We next introduce a dual problem in 
subsection~\ref{sub:Dual} and relate the value function of the original 
firm problem to the value function of the dual problem. Finally, we show 
the operator associated with the dual problem has a unique fixed point (subsection~\ref{sub:uniquedual}). 

\subsection{Primal approach}
\label{sub:Primal}

Here we show by example that the obvious operator associated with the 
original firm problem does not define a contraction mapping. The standard 
approach starts by defining an operator $\sfS$ as follows. For any bounded
function $v$ defined on $\cX$, define the operator by
\begin{eqnarray}
(\sfS v)(x) &=&\sup_{(b^{\prime },d,i)}\big\{d
+\beta
\mathbf{\mathsf{E}}_z v(x')\big\}, \label{prime-operator1}\\
\mbox{subject to}
&&(1-\tau)F(z,k)+\tau\delta k+
[\tau+(1-\tau)q(z,b',k')]b^{\prime}
-b =d+i,\notag\\
&& k'=(1-\delta)k+i,\notag
\\
&& \underbar{$b$}\le b^{\prime }\leq \bar{b},d\geq 0,i\geq 0,\notag\\
&& q(z,b',k')=1/(1+\rho)\mbox{ if }b^{\prime} \leq 0,\notag\\
&& q(z,b',k')=\frac{b^{\prime}\mathbf{\mathsf{E}}%
	_{z}1_{\{v(x^{\prime })> 0\}}+L(k^{\prime })\mathbf{\mathsf{E}}%
	_{z}1_{\{v(x^{\prime })=0\}}}{b^{\prime }\left(1+\rho\right)}\mbox{ if }b'>0,
\label{prime-operator2}
\end{eqnarray}
if the above constraint set is feasible; and $\sfS v=0$ otherwise. A 
Markov-perfect equilibrium is a fixed point of the operator $\sfS$. The usual proof is to show that 
the operator $\sfS$ is a contract mapping under the uniform norm. 
At first it might seem that this would be a contraction mapping
due to the presence of the discounting factor $\beta<1$. However, equality (\ref{prime-operator2}) causes a problem that has not been widely recognized 
in the dynamic corporate finance literature.

The value function $v$ appears in equality (\ref{prime-operator2}) where 
it might not normally appear. Equation (\ref{prime-operator2}) reflects the 
need to take into account potential losses in default and how that affects 
the price of debt. The fact that investors need to account for default 
losses is of course,  well-known. But it creates an extra linkage in the 
model that is not present in the standard Bellman equation setup. In the 
standard model, the objective function takes market prices as given. Here 
the market price $q$ depends on the state dependent $v$ because the 
investor needs to determine the value that will be received in a default.

To provide an example, it is sufficient to find two continuous functions 
$v_1$ and $v_2$ defined on $\cX$ such that $||\sfS v_2-\sfS v_1||\ge ||v_2-v_1||$,
where $||\cdot||$ is a uniform norm defined on $\cX$. Assume $v_1(x)=0$ 
for at least some $x$, so that default may actually happen. We can find a 
simple example with $v_1\equiv 0$.

Assume: 1) there is an exogenous and deterministic shock process, so $\cZ$
contains a single element $z$; 2) set $\tau=0$; 3) set $L(k)=2\nu k^{1/2}$; 4)
set $v_1(x)=0$ and $v_2(x)=\epsilon$ for all $x\in \cX$, where the positive
constants $\nu$ and $\epsilon$ will be chosen. Clearly $||v_2-v_1||=\epsilon$. 
So we need to know about $\sfS v_1$ and $\sfS v_2$.

Part 1. Consider $v_1$, so that
\begin{eqnarray*}
(\sfS v_1)(x) &=&F(z,k)+(1-\delta)k-b +\sup_{k'\ge(1-\delta)k}
\left\{\frac{L(k')}{1+\rho}-k'\right\}\\
&=& F(z,k)+(1-\delta)k-b+\left(\frac{\nu}{1+\rho}\right)^2.
\end{eqnarray*}
Set $k>0$ small enough to ensure that optimal capital is
$k'=\left(\frac{\nu}{1+\rho}\right)^2\ge(1-\delta)k$. Set $b$ small enough to
ensure the above is positive. We now know that the feasible set is non-empty.

Part 2. Consider $v_2$, so that
\begin{eqnarray*}
(\sfS v_2)(x) &=&F(z,k)+(1-\delta)k-b+\beta \epsilon
+\sup_{\underbar{${\scriptstyle b}$}\le b'\le\bar{b},k'\ge(1-\delta)k}
\left\{\frac{b'}{1+\rho}-k'\right\}\\
&=& F(z,k)+(1-\delta)k-b+\beta\epsilon+  \left\{\frac{\bar{b}}{1+\rho}-(1-\delta)k\right\}.
\end{eqnarray*}

Combining Part 1 and Part 2 gives,
\begin{eqnarray*}
(\sfS v_2)(x)-(\sfS v_1)(x)&=&\beta\epsilon+  \left\{\frac{\bar{b}}{1+\rho}-(1-\delta)k\right\}
-\left(\frac{\nu}{1+\rho}\right)^2.
\end{eqnarray*}
Choose $\bar{b}>0$ large enough, and $\epsilon>0$ and $\nu>0$ small enough. Then the
difference can be greater than $\epsilon$. Accordingly we have $||\sfS v_2-\sfS
v_1||>\epsilon=||v_2-v_1||$. This shows that due to default, $\sfS$ is not
necessarily a contraction mapping.

This example shows that the standard approach to defining the operator $\sfS$ 
may not ensure a contraction. Accordingly this direct approach does not work. 
Some other method is needed.

\subsection{Dual problem}
\label{sub:Dual}

In this section, we develop a dual problem to the original problem as done in \cite{aguiar2019contraction}. We start with the following lemma whose proof is in section~\ref{sec-A1}.
\begin{lemma}
	\label{lem-monotone}
	In any equilibrium, $V(z,b,k)$ is strictly decreasing in $b\in
		[\underbar{$b$}(z,k),\bar{b}(z,k)]$. Moreover, $V(z,b_2,k)-V(z,b_1,k)\ge b_1-b_2$
		for any $b_1>b_2$ with $b_1,b_2\in[\underbar{$b$}(z,k),\bar{b}(z,k)]$.
\end{lemma}
Due to the monotonicy, $V(z,b,k)$ has an inverse function with respect to $b$. Let $B(z,k,v)$ be the inverse of the equilibrium value function $V$ in $b$, with 
$v\in [\underbar{$v$}(z,k),\bar{v}(z,k)]$, where $\underbar{$v$}(z,k)=V(z,\underbar{$b$}(z,k),k)$
and $\bar{v}(z,k)=V(z,\bar{b}(z,k),k)$. 
In other words, $V(z,B(z,k,v),k)\equiv v$ for $v\in [\underbar{$v$}(z,k),\bar{v}(z,k)]$ and $B(z,k,V(z,b,k)\equiv b$ for $b\in (\underbar{$b$}(z,k),\bar{b}(z,k))$.

Finding the value function $V$ is the same as finding its inverse, the bond function $B(z,k,v)$. Recall the primal problem that is satisfied by the value function $V$ is a maximization problem (of the value $v$) with given $(z,b,k)$. Next, we construct a dual problem that we will show is satisfied by the bond function $B$. It is a maximization problem (of the bond $b$) with given $(z,k,v)$.

To specify the dual problem, we introduce a three-dimensional dual
space $\cY=\cZ\times \Re_+^2$. For $y=(z,k,v)\in\cY$, the first component $z$
is the exogenous state, the second component $k$ is the capital level, and the
third component $v$ is the value of the firm.

\smallskip \textit{The dual}. Let $x'=(z',b',k')$. Given an equilibrium condition, we
define the dual problem for (\ref{prime-operator1}) as follows:
\begin{eqnarray}
&&\hat{B}(z,k,v)=\sup_{(b^{\prime },d,i)}\Big\{
(1-\tau)F(z,k)+\tau\delta k-d-i+\tau b'\notag\\
&&\phantom{\hat{B}(z,k,v)=}\quad
+\frac{1-\tau}{1+\rho}
\big\{b' 1_{\{b'\le0\}}+\big[b^{\prime}\mathbf{\mathsf{E}}_{z}1_{\{V(x^{\prime })> 0\}}
+L(k^{\prime })\mathbf{\mathsf{E}}_{z}1_{\{V(x^{\prime })=0\}}\big]1_{\{b'>0\}}\big\}\Big\}
\label{eq:FirmBellman-s-dual}\\
&&\mbox{subject to}  \notag \\
&&\qquad (1-\tau)F(z,k)+\tau\delta k-d-i+\tau b'\notag\\
&&\qquad\qquad\qquad
+\frac{1-\tau}{1+\rho}
\big\{b' 1_{\{b'\le0\}}+\big[b^{\prime}\mathbf{\mathsf{E}}_{z}1_{\{V(x^{\prime })> 0\}}
+L(k^{\prime })\mathbf{\mathsf{E}}_{z}1_{\{V(x^{\prime })=0\}}\big]1_{\{b'>0\}}\big\}\Big\}\ge0,\notag\\
&&\qquad v=d+\beta \mathbf{\mathsf{E}}_z  V(x'),\notag \\
&&\qquad k'=(1-\delta)k+i,\notag
\\
&&\qquad \underbar{$b$}\le b^{\prime }\leq \bar{b},d\geq 0,i\geq 0.\notag
\end{eqnarray}
The following lemma confirms that $\hat{B}(z,k,v)$ in the dual problem
(\ref{eq:FirmBellman-s-dual}) indeed gives the inverse function $B(z,k,v)$.

\begin{lemma}
	\label{lem-dual-equiv}
	$B(z,k,v)=\hat{B}(z,k,v)$.
\end{lemma}
For the proof see section \ref{proof:lem-dual-equiv}.

The next step is to define an operator and show that $B$ is a fixed point of
this operator. Define an operator $\sfT$ on all continuous function $f$ of
$y=(z,k,v)\in\cY$ as follows,
\begin{eqnarray}
&& (\sfT f)(z,k,v)=\sup_{(b^{\prime },d,i)}\Big\{
(1-\tau)F(z,k)+\tau\delta k-d-i+\tau b'\notag\\
&&\phantom{\sfT f(z,k,v)=}\quad
+\frac{1-\tau}{1+\rho}
\big\{b' 1_{\{b'\le0\}}+\big[b^{\prime}\mathbf{\mathsf{E}}_{z}1_{\{w(x^{\prime })> 0\}}
+L(k^{\prime })\mathbf{\mathsf{E}}_{z}1_{\{w(x^{\prime })=0\}}\big]1_{\{b'>0\}}\big\}\Big\}
\label{eq:dual-operator}\\
&&\mbox{subject to}  \notag \\
&&\qquad (1-\tau)F(z,k)+\tau\delta k-d-i+\tau b'\notag\\
&&\qquad\qquad\qquad
+\frac{1-\tau}{1+\rho}
\Big\{b' 1_{\{b'\le0\}}+\big[b^{\prime}\mathbf{\mathsf{E}}_{z}1_{\{w(x^{\prime })> 0\}}
+L(k^{\prime })\mathbf{\mathsf{E}}_{z}1_{\{w(x^{\prime })=0\}}\big]1_{\{b'>0\}}\Big\}
\ge0, \notag\\
&&\qquad v=d+\beta \mathbf{\mathsf{E}}_z  w(x'),\notag \\
&&\qquad b'\le f(z',k',w(x'))\mbox{ for all }x'\in\cX\mbox{ such that }w(x')\ge0
\mbox{ for all }x'\in\cX,\notag\\
&&\qquad k'=(1-\delta)k+i,\notag
\\
&&\qquad \underbar{$b$}\le b^{\prime }\leq \bar{b},d\geq 0,i\geq 0,\notag
\end{eqnarray}%
where $q^*=1/(1+\rho)$.

The maximization problem (\ref{eq:FirmBellman-s-dual}) is to maximize the bond $b$. So we expect any feasible bond $b\le \hat{B}(z,k,v)=B(z,k,v)$. This implies that $b'\le B(z',k',v')$. By constructing the operator above, the goal is to show that $B$ is a fixed point. Then for any continuous function $f$, we require $b'\le f(z',k',v')$, with the next period value function constructed by $v'=w(x')$. The next lemma confirms that $B$ is in fact a fixed point of the constructed operator.

\begin{lemma}
	\label{lem-BisFixedPT}
	Any equilibrium $B(z,k,v)$ is a fixed point of $\sfT$.
\end{lemma}
For the proof see section \ref{proof:lem-BisFixedPT}.

With this lemma it is sufficient for us to prove that the operator $\sfT$ has a
unique fixed point. 

\subsection{Unique fixed point for the dual operator}
\label{sub:uniquedual}

Typically, the operator takes the form
$\sfT v=sup \{r +\beta E(v)\}$ subject to some constraints. We need to be sure
that $r$ is bounded, otherwise it is not clear that the operator  $\sfT$ is
bounded and discounting.
So we cannot just directly apply \cite{blackwell1965discounted}.

Some papers directly add an extra assumption saying that $r$ is bounded.
Another approach is to use other features of the model to identify a bound. For
example, \cite{harris1987dynamic} studies an optimal growth model. He obtains
a bound inductively primarily by using the flow resource constraints.
Either of these could potentially be applied to our problem.

We have instead chosen to follow \cite{lippman1975dynamic} and
\cite{nunen1978}. To do that we identify a function $\phi$ in Lemma
\ref{lem-r-bound}. This function ensures that for any bounded continuous
function $f$: 1) $\sfT$ satisfies the  monotonicity property (Lemma
\ref{lem-T-monotone}), 2) $\sfT f$ is bounded under $\phi$-norm (Lemma
\ref{lem-T-bounded}), and, 3) it satisfies the discounting property
(Lemma \ref{lem-T-discounting}). Then the existence and uniqueness of a fixed point
for $\sfT$ is concluded by applying
\cite{blackwell1965discounted} as in \cite{nunen1978}.

The first step is monotonicity, which clearly holds.
\begin{lemma}
	\label{lem-T-monotone}
	(Monotonicity) For any bounded continuous functions $f$ and $g$ on $\cY$,
	if $f\le g$, then $\sfT f\le \sfT g$.
\end{lemma}

As a preparation for the next two steps, we define the norm. For 
boundedness and discounting we need to identify a positive function 
$\phi$. Then we need to show that the operator $\sfT$ is bounded and 
satisfies discounting, under the $\phi$-norm. For any positive function 
$\phi(y)$ and for any $y=(z,k,v)\in\cY$, we define $\phi$-norm of function 
$f$ on $\cY$ as
\begin{eqnarray}
	||f||_\phi=\sup_{y} \left\{\frac{|f(y)|}{\phi(y)}:y\in\cY\right\}.
\end{eqnarray}

We choose the function $\phi$ for the operator $\sfT$ as follows. Let
$\cA(z,k,v)$ denote the set of the feasible policy for the operator problem
(\ref{eq:dual-operator}); that is, the set of all $(b',d,i)$ that satisfy
the constraints of the optimization problem (\ref{eq:dual-operator}).

The second step is to establish boundedness. To do this we write the single
period objective function as follows. For any $y=(z,k,v)\in\cY$ and
$a=(b',d,i)\in \cA(z,k,v)$, the objective function in (\ref{eq:dual-operator})
can be written as
\begin{eqnarray*}
	r(y,a)&=&(1-\tau)F(z,k)+\tau\delta k-d-i+\tau b'\notag\\
	&&+\frac{1-\tau}{1+\rho}
\Big\{b' 1_{\{b'\le0\}}+\big[b^{\prime}\mathbf{\mathsf{E}}_{z}1_{\{w(x^{\prime })> 0\}}
+L(k^{\prime })\mathbf{\mathsf{E}}_{z}1_{\{w(x^{\prime })=0\}}\big]1_{\{b'>0\}}\Big\},
\end{eqnarray*}
where the state $x'=(z',b',k')$. We call $r(y,a)$ the single-period reward
function.

\begin{lemma}
	\label{lem-r-bound}
	For any $\epsilon>0$, there exists an  $\eta>0$ such that the function,
	\begin{eqnarray*}
\phi(y)&\equiv&		
\phi(z,k,v)\equiv \phi_0(k) \\
&=&\left[(1-\tau)\bar{z}A+\tau\delta\right]k+\eta,
	\end{eqnarray*}
	ensures that for all $y\in\cY$,
	\begin{eqnarray*}
		r(y,a)\le \phi(y)\mbox{ for all }a\in \cA(y)\qquad\mbox{and}
		\qquad \phi(y') \le (1+\epsilon) \phi(y) \mbox{ for all } a\in \cA(y),
	\end{eqnarray*}
	where $\phi(y')\equiv \phi_0(k')$ with $k'=(1-\delta)k+i$.
\end{lemma}
For the proof see section \ref{proof:lem-r-bound}.

With this choice of $\phi$ function, we have
\begin{lemma}
	\label{lem-T-bounded}
	(Boundedness) Let $f$ be a bounded continuous function on $\cY$. Then
	$\sfT f$ is bounded under $\phi$-norm.
\end{lemma}

This lemma follows immediately from Lemma~\ref{lem-r-bound}. Note that even if
$f$ is bounded, $\sfT f$ may not be bounded under the uniform norm, which is 
needed in order to apply \cite{blackwell1965discounted}.

The third step is to establish discounting.
\begin{lemma}
	\label{lem-T-discounting}
	(Discounting)
	For any bounded continuous function $f$ on $\cY$ and constant $a>0$,
	$\sfT (f+a\phi)\le \sfT f+\theta a\phi$ for some $\theta<1$.
\end{lemma}
For the proof see section \ref{proof:lem-T-discounting}.

With monotonicity, boundedness, discounting all established, we can now apply
the \cite{blackwell1965discounted} sufficient conditions for a contraction to
the dual. Accordingly, there exists a unique Markov perfect equilibrium.

Because the dual problem satisfies the conditions for a contraction mapping, 
using value function iteration on the dual to find the unique solution is 
justified. So there may be an advantage for people working on numerical 
solutions to such models, to work with the dual problem.

To summarize: the model has a financial problem and a technical problem. The
financial problem reflects the fact that investors must expect to break-even.
The technical problem reflects the need to establish the existence of a 
unique equilibrium. Both of these issues can be addressed in several ways, 
and the best method may depend to some degree on the specifics of a model. 
We believe that the approach taken here may be useful for a variety of 
applications since break-even requirements for outside participants are 
common to many models of firms.

\section{Equilibrium capital, debt and dividends}
\label{sec-capital-debt-dividend}

From Theorem \ref{thm-main-theorem} we know that there exists a unique
equilibrium consisting of a value function $V$ and price function $q$. 
What can be said about the properties of the equilibrium? In this section, 
we consider a number of properties regarding equilibrium capital, debt, 
and dividends. Throughout the rest of this subsection, we assume 
$x=(z,b,k)\in\cX_{feas}$. Otherwise, the firm would default immediately 
which is not so interesting.

We start with some properties of the value function, the price of 
the debt, and the debt level. Specifically,  the value function is shown to strictly 
decrease with the debt level. The price of the debt is shown to decrease 
in the amount of debt issued. The debt level is shown to be nonnegative; that is, the firm does not save. The proof of this proposition is in section \ref{proof:sec-capital-debt-dividend}.

\begin{proposition}
	\label{pro-monotone}
	In the equilibrium,
	\begin{description}
		\item[(a)] $V(z,b,k)$ is strictly decreasing in $b\in
		[\underbar{$b$}(z,k),\bar{b}(z,k)]$. Moreover, $V(z,b_2,k)-V(z,b_1,k)\ge b_1-b_2$
		for any $b_1>b_2$ with $b_1,b_2\in[\underbar{$b$}(z,k),\bar{b}(z,k)]$.
		\item[(b)]  $q(z,b',k')$ is decreasing in $b'$. 
\item[(c)] $b'\ge0$, so there is no saving by the firm.
	\end{description}
\end{proposition}

To go further it is useful to distinguish three cases depending on the current
state $x=(z,b,k)$. We refer to these as high capital, moderate capital, and low
capital, and they are defined by relative to the ability to reach the capital 
target $k^*(z)$, which depends on the current exogenous state as 
determined by the equality (\ref{eq-optimaltarget}).

In the first case (``high capital''), the firm already has a great deal of 
capital. The current capital level is high enough that the firm does not want 
to invest in more. The firm will optimize the debt level under the current 
state $z$ and the future capital level $(1-\delta)k$. The excess is paid out 
as dividends. Whether the state is high enough for this to happen depends on 
whether depreciated capital $(1-\delta)k$ is larger than a target capital 
level $k^*(z)$. 

In the second case (``moderate capital''), the capital target is attainable. 
Current debt is not too high and the capital level is not too low. Their 
combination allows the firm to raise enough cash through production and 
optimal debt. The firm's capital is raised to the optimal target for next 
period. Again, the firm will pay the excess funds as dividends. Proposition~\ref{pro-moderate-capital} provides a formal characterization.

In the third case (``low capital''), the capital target is not attainable. 
Given current debt and capital, even with the best feasible capital and debt 
choices, the firm cannot reach the optimal target level. The firm will not 
pay current dividends due to the great need for capital investment.

We have not been able to completely prove the high capital and low capital 
cases formally. We provide a more detailed description as conjectured. 

To describe the target, define
\begin{eqnarray}
	N(z,k')&=&\max_{0\le b'\le\bar{b}} \big[\tau+(1-\tau)q(z,b',k')-\beta\big] b',
\label{eq-optimal-Nchoice}\\
	b^*(z,k')&=&\arg\max_{0\le b'\le\bar{b}}
	\big[\tau+(1-\tau)q(z,b',k')-\beta\big] b'.
\label{eq-optimal-debt1}
\end{eqnarray}
$N(z,k')$ is the maximum revenue from the bond issuing, and $b^*(z,k')$ is 
the optimal bond level from issuing under the current exogenous state $z$ and 
the future capital level $k'$. Notice that we have replaced lower bound 
$\underbar{$b$}$ by $0$ in the above due to 
Proposition~\ref{pro-monotone} (c).

The optimal capital target is determined by
\begin{eqnarray}
k^*(z)=\arg\max_{k'}\Big\{ \beta(1-\tau)\ex_z F(z',k')
-\big[1-\beta+\beta\delta(1-\tau)\big] k'+N(z,k')\Big\}.\label{eq-optimaltarget}
\end{eqnarray}
Corresponding to this capital target, there is a debt target $b^*(z)\equiv b^*(z,k^*(z))$
(where $b^*(z,k)$ is determined by (\ref{eq-optimal-debt1}))
and a dividend target
\begin{eqnarray}
d^*(z)=R(z,k)+N\left(z, k^*(z)\right)+\beta b^*(z)-k^*(z)-b.\label{eq-d-star}
\end{eqnarray}

\begin{proposition}
\label{pro-moderate-capital}
Suppose that $x=(z,b,k)\in\cX_{feas}$. If $k^*(z)\ge(1-\delta)k$ and
$d^*(z)\ge0$, then the equilibrium optimal investment $i=k^*(z)-(1-\delta)k$
(or equivalently capital level is raised to $k^*(z)$), optimal debt
$b'=b^*(z)$, and optimal dividend $d=d^*(z)$.
\end{proposition}
For the proof see section \ref{proof:sec-capital-debt-dividend}.

The conditions $k^*(z)\ge(1-\delta)k$ and $d^*(z)\ge0$ in the above proposition
ensure that the constraints related to $i\ge0$ and $d\ge0$ in the firm problem
(\ref{eq:FirmBellman-s}) are satisfied automatically. When the condition $k^*(z)\ge(1-\delta)k$ fails, the constraint $i\ge0$ must be explicitly imposed
in the firm's problem. This corresponds the high capital case. When the condition $k^*(z)\ge(1-\delta)k$ holds but the condition $d^*(z)<0$, the constraint $d\ge0$ must be explicitly imposed in the firm's problem. This corresponds the low capital level case.

Now we provide a more detailed description of the high capital level case, with $(1-\delta)k>k^*(z)$. The current capital level after depreciation is already larger than the optimal target. So in order to achieve this target, the firm would need to make a negative investment $i=k^*(z)-(1-\delta)k<0$. But negative investment is not allowed by assumption. The closest solution would be a zero investment. So we conjecture that the equilibrium optimal investment is  $i=0$. This means that the next period capital is $k'=(1-\delta)k$.
Mathematically, this conjecture says that $k'=(1-\delta)k$ is the optimal solution to the maximization problem (\ref{eq-optimaltarget}) subject to the constraint $k'\ge(1-\delta)k$. A sufficient condition is that the maximand in (\ref{eq-optimaltarget}) is quasi-concave in $k'$. The challenge
to establish this lies in the difficulty in characterizing the function $N(z,k')$, which is defined by (\ref{eq-optimal-Nchoice}) and which involves the equilibrium price function $q(z,b',k')$.

Corresponding to this optimal capital level, the optimal debt
$b'=b^*(z,(1-\delta)k)$ and the optimal dividend (which maximizes
(\ref{eq-optimal-Nchoice}))
\begin{eqnarray}
d^*_1(z)&=&(1-\tau)F(z,k)+\tau\delta k +N\left(z,(1-\delta)k\right)
+\beta b^*(z,(1-\delta)k)-b.
\end{eqnarray}
The feasibility assumption
$x=(z,b,k)\in\cX_{feas}$ guarantees $d^*_1(z)\ge0$.

Finally consider the low capital case. So, $k^*(z)\ge(1-\delta)k$ and
$d^*(z)<0$. Dividends cannot be negative, so the constraint $d\ge0$ in the firm problem (\ref{eq:FirmBellman-s}) may be binding; and we conjecture that it is binding. It is optimal not to pay any dividends if the firm has inadequate capital. The firm  currently has a small amount of capital and a high debt. It cannot raise enough total revenue from production and new debt to payoff the existing debt and at the same time raise the capital level to the target $k^*(z)$. However, recall that we are assuming $x\in\cX_{feas}$, so the firm can raise enough to avoid default. Under the conjecture, the equilibrium optimal debt and capital pair $(b',k')$ is the solution to the following optimization problem,
		\begin{eqnarray*}
			\max_{0\le b'\le \bar{b}\atop k'\ge(1-\delta)k} && \ex_z V(z',b',k')\\
			\mbox{s.t.} && (1-\tau)F(z,k)+[1-\delta(1-\tau)]k+\left[\tau+(1-\tau)
			q(z,b',k')\right] b'-k'=b,
		\end{eqnarray*}
where the last equality is the budget flow equation with $d=0$.

Financial targets play a prominent role in the capital structure literature at least since \cite{myers1984capital}; see \cite{deangelo2015stable}, \cite{faulkender2012cash}, \cite{frank2019corporate} and \cite{ai2021trade}. Some papers study time invariant targets. It is important to stress that here the target is state dependent.

We close this section with a comparison to the optimal target of a financial autarky model which is a natural benchmark. In this model, the firm can choose capital, but there is no debt market. Otherwise, the problem has the same structure as above. Any capital purchase must be paid for using revenue generated from  operations. Of course, if the firm has no capital to start with it can never purchase any. So we assume that the firm has some capital to start with.
It is shown in \cite{chenfrank2021} that the optimal capital target is
\begin{eqnarray}
	k^*_n(z)=\arg\max_{k'}\Big\{ \beta(1-\tau)\ex_z F(z',k')
	-\big[1-\beta+\beta\delta(1-\tau)\big] k'\Big\}. \label{eq-optimaltarget-Autarky}
\end{eqnarray}
This target has an explicit solution,
\begin{equation*}
	k^*_n(z)= \left(\frac{\alpha\beta A (1-\tau)\ex_z\left(z'\right)}{1-\beta
		+\delta\beta(1-\tau)}\right)^{1/(1-\alpha)}.
\end{equation*}
The difference between the target $k^*_n(z)$ in equation
(\ref{eq-optimaltarget-Autarky}) and the target $k^*(z)$ in equation
(\ref{eq-optimaltarget})  is
the extra term $N(z,k')$ in (\ref{eq-optimaltarget}). Since the maximand in
equation (\ref{eq-optimaltarget-Autarky}) is concave in $k'$ and $N(z,k')$ is
non-decreasing in $k'$, we know that $k^*(z)\ge k^*_n(z)$. In other words, the
target capital with a debt market is greater than the target capital when there
is no debt market.

\section{Model modifications}
\label{sec:extend}

We have provided a method to establish existence and uniqueness for a very
standard trade-off model. Our approach seems applicable to many of the cases
studied in the literature. To help show this, in this section we consider two
popular alternative assumptions to those in our basic model.

\subsection{Exogenous cash flow}
Starting with \cite{modigliani1958cost}, many capital structure papers have
taken the firm's cash flows to be exogenous, see \cite{leland1994corporate,goldstein2001ebit,abel2018optimal,demarzo2021}. The
original motivation was to remove the impact of the firm's real investment decisions on the firm's choice of debt. Of course, as shown in our basic model these are fundamentally connected due to the terms on which investors will provide funds. Here we show that removing that link simplifies the model and our approach still works.

Assume that the cash flows follow an exogenous process $\{z_t,t\ge0\}$. The
state space $\cX=\cZ\times [\underbar{$b$},\bar{b}]$. The firm flow budget
(\ref{eq:FirmBudget-borrow}) is modified to
\begin{equation}
z_t+\tau[1-q(z_t,b_{t+1})]b_{t+1}
+q(z_t,b_{t+1})b_{t+1}-b_{t}
=d_{t}+ i_{t}.  \label{eq:FirmBudget-borrow-exogen}
\end{equation}
Let
\begin{eqnarray}
S(z)&=&\sup_{\underbar{$b$}\le b'\le \bar{b}}[\tau+(1-\tau)q(z,b')]b^{\prime}.
\label{debt-prod-B-exogen}
\end{eqnarray}
The maximum possible net revenue of the firm is given by $z+S(z)-b$, which must
be non-negative for the firm to avoid default. The feasible set in any equilibrium is defined as, $ \mathbb{X}_{feas}= \Big\{(z,b)\in\cX\ \big| \
z + S(z) \ge b\Big\}$. Let $\underbar{$b$}(z)=\inf\{b\ge\underbar{$b$}:(z,b)\in
\mathbb{X}_{feas}\}$ and $\bar{b}(z)=\sup\{b\le\bar{b}:(z,b)\in \mathbb{X}_{feas}\}$.

\smallskip\noindent
\textit{Firm problem.}
Let $x=(z,b)\in\cX$ and take $q=q(z,b')$ as given. If $x\in \mathbb{X}_{feas}$, then
\begin{eqnarray}
\label{eq:FirmBellman-s-exogen}
V(x)&=&\sup_{\underbar{${\scriptstyle b}$}\le b^{\prime }\le\bar{b},d\ge0}\quad d +\beta
\mathbf{\mathsf{E}}_z V(x'), \\
&& \mbox{subject to}  \quad z+[\tau+(1-\tau)q(z,b')]b^{\prime}
-b =d,\notag
\end{eqnarray}%
otherwise, $V(x)=0$.

\smallskip\noindent
\textit{Investor break-even condition.} Given the firm value function $V$ and
the firm's exogenous state $z$. If the bond issue is $b^{\prime} \leq 0$, then $q(z,b')=1/(1+\rho)$. If $b^{\prime } \geq 0$, then
\begin{align}
q(z,b')&=\frac{\mathbf{\mathsf{E}}%
	_{z}1_{\{V(x^{\prime })> 0\}}}{1+\rho}.\label{eq-upperbound-q-exogen}
\end{align}

Similar to the main model (Definition~\ref{Equilibrium-definition}), a
Markov-perfect equilibrium is defined to be a value function $V$ and a price
function $q$ such that
\begin{description}
	\item[(1)] Given $q$,  the value function $V$ satisfies the Bellman
	equation (\ref{eq:FirmBellman-s-exogen}), and the supremum in
	(\ref{eq:FirmBellman-s-exogen}) is attained for some policy;
	\item[(2)] Given $V$, $q$ satisfies the break-even condition
	(\ref{eq-upperbound-q-exogen}).
\end{description}

\begin{proposition}
\label{thm-main-proposition}
Under Assumption~\ref{ass-bounded-shock}, there exists a unique Markov-perfect
equilibrium.
\end{proposition}
The proof is provided in Section \ref{proof:sec:extend}.

Exogenous cash flow capital structure models and investment based capital
structure models are often treated as if they are entirely distinct approaches,
see \cite{strebulaev2012dynamic} and \cite{ai2021trade}. In practice, it is
common to use continuous time for models with exogenous cash flows and  discrete time for models with an investment choice. However, the models are not as different as they sometimes appear. The fact that the same approach works in a simpler model should not be surprising. The two approaches really are closely connected.

\subsection{Capital adjustment costs}

Both the investment literature \citep{hayashi1982tobin,cooper2006nature} and 
the dynamic capital structure literature \citep{strebulaev2012dynamic} commonly 
assume that
investment is subject to adjustment costs. A quadratic functional form is
particularly popular, as in \cite{nikolov2021sources}. With such well behaved adjustment costs our approach still applies. To see this assume that capital adjustment costs are,
\begin{equation}
\Psi(k_t,k_{t+1})=\frac12\psi\left(\frac{k_{t+1}-(1-\delta)k_t}{k_t}\right)^2k_t
\equiv \frac12\psi\left(\frac{i_t}{k_t}\right)^2k_t.
\end{equation}
The firm flow budget condition (\ref{eq:FirmBudget-borrow}) becomes
\begin{equation*}
\underbrace{(1-\tau)F(z_{t},k_{t})}_{\substack{ \text{revenue}  \\
		\text{after-tax}}}+\underbrace{\tau (\delta
		k_{t}+(1-q_{t})b_{t+1})}_{%
	\text{tax deduction revenue}}
+\underbrace{q_{t}b_{t+1}-b_{t}}_{\substack{ \text{net debt}  \\
		\text{revenue}}}=\underbrace{d_{t}}_{\text{dividends}}+\underbrace{i_{t}%
}_{\text{investment}}+\underbrace{\Psi(k_t,k_{t+1})}_{\substack{ \text{adjustment}  \\
		\text{cost}}}.
\end{equation*}
The definition of the function $S$ in (\ref{debt-prod-B}) becomes,
\begin{eqnarray}
S(z,k)&=&\sup_{\underbar{${\scriptstyle b}$}\le b'\le \bar{b}
\atop
k'\ge(1-\delta)k}\big\{[\tau+(1-\tau)q(z,b',k')]b^{\prime}-k'-\Psi(k,k')\big\}.
\label{debt-prod-B-adjustment}
\end{eqnarray}
The definition~\ref{def-feasible} for the feasible set remains the same.
The firm problem remains almost the same. The change is that the first
constraint in the Bellman optimization (\ref{eq:FirmBellman-s}) is replaced by
\begin{eqnarray*}
R(z,k)+[\tau+(1-\tau)q(z,b',k')]b^{\prime} -b =d+k'+\Psi(k,k').
\end{eqnarray*}%

The existence and uniqueness of a Markov perfect equilibrium still applies. The
proof only requires minor adjustments to the proof of Theorem \ref{thm-main-theorem}. Quadratic capital adjustment costs do not fundamentally disrupt the proof.

\section{Conclusion}
\label{sec-conclusion}

If outside investors are going to fund a firm, they must expect to break-even. 
Otherwise, they will not participate. The price of debt adjusts accordingly. This
basic finance requirement means that even a very basic trade-off theory model with 
single period risky debt may not directly satisfy the usual 
\cite{stokey1989recursive} method of proving existence and uniqueness.

We first illustrated the difficulty. Then we showed how to establish existence and uniqueness. This was done by turning to the dual problem, and then changing the measure. We provide two further closely related versions of the model where the approach also applies. So this approach may be useful for a range of such models.

A noteworthy feature of the model is that in the equilibrium, the firm has a
state dependent capital target and a debt target. The firm's debt and dividends 
decisions adapt to the needs of productive efficiency. While this description 
sounds reminiscent of the pecking order \citep{myers1984capital}, the underlying 
mechanism is of course quite different. There is no adverse selection in our model. 
Productive efficiency subject to feasibility constraints, is the key driving force. 
We also show that when the firm has access to a debt market the capital target is 
higher than when the firm is required to self-finance.

\appendix

\section{Proofs of lemmas in section~\ref{sec-proof}}
\label{sec-A1}

\proof{Proof of Lemma~\ref{lem-monotone}}
In this proof, we assume that all the states are in $\cX_{feas}$. 
Fix $(z,k)$ with $z\in\cZ$ and $k\ge0$.
	For any given $b_1,b_2\in[\underbar{$b$}(z,k),\bar{b}(z,k)]$ with $b_1>b_2$,
	choose  any $0<\epsilon<b_1-b_2$. Following the definition of the firm's
	problem
	(\ref{eq:FirmBellman-s}), there exists a policy $(b'_1,d_1,i_1)$ such that
	\begin{eqnarray*}
		&&V(z,b_1,k)<d_1+\beta\ex_z V(z',b'_1,k'_1)-\epsilon, \mbox{and}\\
		&&(1-\tau)F(z,k)+\tau\delta k+
		[\tau+(1-\tau)q(z,b'_1,k'_1)]b^{\prime}_1
		-b_1 =d_1+i_1,
	\end{eqnarray*}
	where $k'_1=(1-\delta)k+i'_1$.
	
	If we replace $b_1$ by $b_2$ and $d_1$ by $d_2=d_1+b_1-b_2$, then the
	policy $(b'_1,d_2,i_1)$ is feasible with the given state $(z,b_2,k)$. In
	particular,	the budget constraint still holds and the future state remains
	the same ($b'_2=b'_1$ and $k'_2=k'_1$). So the above inequality can be
	written as
	\begin{eqnarray*}
		V(z,b_1,k)&<&d_1+\beta\ex_z V(z',b'_1,k'_1)-\epsilon\\
		&=&d_2+\beta\ex_z V(z',b'_1,k'_1)-(b_1-b_2)-\epsilon  \le
		V(z,b_2,k)-(b_1-b_2)-\epsilon.
	\end{eqnarray*}
	Because $\epsilon>0$ can be arbitrarily small, the above inequality implies
	\begin{eqnarray*}
		V(z,b_1,k)\le V(z,b_2,k)-(b_1-b_2).
	\end{eqnarray*}
\endproof

\proof{Proof of Lemma~\ref{lem-dual-equiv}.}
\label{proof:lem-dual-equiv}
	Fix $x_0=(z_0,b_0,k_0)\in\cX_{feas}$. Suppose that $(b'_0,d_0,i_0)$ is a feasible solution to the
	equilibrium problem (\ref{prime-operator1}). Let $k'_0=(1-\delta)k_0+i_0$,
	$v_0=V(x_0)$, $x'_0=(z'_0,b'_0,k'_0)$, and
	\begin{eqnarray*}
		d_0&=& (1-\tau)F(z_0,k_0)+\tau\delta k_0-b_0-i_0 +\tau b'_0
\notag\\
		&&+\frac{1-\tau}{1+\rho}
\Big\{b'_0 1_{\{b'_0\le0\}}+\big[b^{\prime}_0\mathbf{\mathsf{E}}_{z_0}1_{\{V(x^{\prime }_0)> 0\}}
+L(k^{\prime }_0)\mathbf{\mathsf{E}}_{z_0}1_{\{V(x^{\prime }_0)=0\}}\big]1_{\{b'_0>0\}}\Big\}.
	\end{eqnarray*}
	Now it is immediate to verify that with $(z_0,k_0,v_0)$, the policy $(b'_0,d_0,i_0)$ is
	feasible for the dual problem (\ref{eq:FirmBellman-s-dual}); hence,
	\begin{eqnarray*}
		b_0&=& (1-\tau)F(z_0,k_0)+\tau\delta k_0-d_0-i_0 +\tau b'_0
\notag\\
		&&+\frac{1-\tau}{1+\rho}
\Big\{b'_0 1_{\{b'_0\le0\}}+\big[b^{\prime}_0\mathbf{\mathsf{E}}_{z_0}1_{\{V(x^{\prime }_0)> 0\}}
+L(k^{\prime }_0)\mathbf{\mathsf{E}}_{z_0}1_{\{V(x^{\prime }_0)=0\}}\big]1_{\{b'_0>0\}}\Big\}\\
		&\le &\hat{B}(z_0,k_0,v_0).
	\end{eqnarray*}
	Since $B(z,k,v)$ is the inverse function of $V(z,\cdot,k)$, we have
	$b_0=B(z_0,k_0,v_0)$. Hence, we have shown $B(z_0,k_0,v_0)\le \hat{B}(z_0,k_0,v_0)$.
	
	Suppose that $B(z_0,k_0,v_0) < \hat{B}(z_0,k_0,v_0)$.
	For any $\epsilon>0$, there exists a feasible policy
	$(\hat{b}'_\epsilon,\hat{d}_\epsilon,\hat{i}_\epsilon)$
	for the dual problem  (\ref{eq:FirmBellman-s-dual}) such that
	\begin{eqnarray*}
		\hat{B}(z_0,k_0,v_0)-\epsilon
		&<&(1-\tau)F(z_0,k_0)+\tau\delta k_0-\hat{d}_\epsilon
		-\hat{i}_\epsilon+\tau \hat{b}'_\epsilon \notag\\
		&&+\frac{1-\tau}{1+\rho}
\Big\{\hat{b}'_\epsilon 1_{\{\hat{b}'_\epsilon \le 0\}}
+\big[\hat{b}^{\prime}_\epsilon
			\mathbf{\mathsf{E}}%
			_{z_0}1_{\{V(\hat{x}'_\epsilon )> 0\}}+L(k^{\prime })\mathbf{\mathsf{E}}%
			_{z_0}1_{\{V(\hat{x}'_\epsilon )=0\}}\big]
1_{\{\hat{b}'_\epsilon >0\}}\Big\},
	\end{eqnarray*}
	where $\hat{k}'_\epsilon =
	(1-\delta)k_0+\hat{i}_\epsilon $ and $\hat{x}'_\epsilon =(z'_0,\hat{b}'_\epsilon ,\hat{k}'_\epsilon )$.
	Let $\hat{b}_0=\hat{B}(z_0,k_0,v_0)$ and note $v_0=\hat{d}_\epsilon
	+\beta\mathbf{\mathsf{E}}_{z_0}V(\hat{s}'_\epsilon)$. We can rewrite the above inequality
	as
	\begin{eqnarray*}
		v_0-\epsilon
		&<&(1-\tau)F(z_0,k_0)+\tau\delta k_0
		-\hat{i}_\epsilon+\tau \hat{b}'_\epsilon \notag\\
		&&+\frac{1-\tau}{1+\rho}
\Big\{\hat{b}'_\epsilon 1_{\{\hat{b}'_\epsilon \le 0\}}
+\big[\hat{b}^{\prime}_\epsilon
			\mathbf{\mathsf{E}}%
			_{z_0}1_{\{V(\hat{x}'_\epsilon )> 0\}}+L(k^{\prime })\mathbf{\mathsf{E}}%
			_{z_0}1_{\{V(\hat{x}'_\epsilon )=0\}}\big]
1_{\{\hat{b}'_\epsilon >0\}}\Big\}
+\beta\mathbf{\mathsf{E}}_{z_0}V(\hat{x}'_\epsilon)\\
		&\le& V(z_0,\hat{b}_0,k_0),
	\end{eqnarray*}
	where the last inequality is due to the fact that
	the policy $(\hat{b}'_\epsilon,\hat{d}_\epsilon,\hat{i}_\epsilon)$ is a feasible policy for the
	equilibrium problem (\ref{prime-operator1}) with $(z_0,\hat{b}_0,k_0)$.
	Hence, we show that $v_0-\epsilon = V(z_0,b_0,k_0)-\epsilon< V(z_0,\hat{b}_0,k_0)$ for
	all $\epsilon>0$. Let $\epsilon\to 0$; we have
	$V(z_0,b_0,k_0)\le V(z_0,\hat{b}_0,k_0)$. Since $V(z,b,k)$ is monotone decreasing in $b$,
	we have $b_0\ge \hat{b}_0$ or
	$B(z_0,k_0,v_0) \ge \hat{B}(z_0,k_0,v_0)$, a contradiction.
	\hfill $\Box$
\endproof

\proof{Proof of Lemma~\ref{lem-BisFixedPT}.}
\label{proof:lem-BisFixedPT}
	Fix any $x_0=(z_0,k_0,v_0)\in\cY$. We first show that $B(z_0,k_0,v_0)\le
	(\sfT B)(z_0,k_0,v_0)$. For any $\epsilon>0$,
	there exists a feasible policy $(b^{\prime }_0,d_0,i_0)$ to the dual problem
	(\ref{eq:FirmBellman-s-dual}) such that
	\begin{eqnarray*}
		B(z_0,k_0,v_0)&<&
		(1-\tau)F(z_0,k_0)+\tau\delta k_0-d_0-i_0+\tau b'_0\notag\\
		&&+\frac{1-\tau}{1+\rho}
\Big\{b'_0 1_{\{b'_0\le0\}}+\big[b^{\prime}_0\mathbf{\mathsf{E}}_{z_0}1_{\{V(x^{\prime }_0)> 0\}}
+L(k^{\prime }_0)\mathbf{\mathsf{E}}_{z_0}1_{\{V(x^{\prime }_0)=0\}}\big]1_{\{b'_0>0\}}\Big\}
+\epsilon.
	\end{eqnarray*}
	Let $w(x'_0)=V(x'_0)$. Then, $b'_0=B(z'_0,k'_0,V(x'_0))
	=B(z'_0,k'_0,w(x'_0))$ as $B$ is an inverse function of $V$ in terms of its second argument.
	This shows that $(b^{\prime }_0,d_0,i_0)$ and $w(x'_0)$ is feasible to the $\sup$ problem
	in the definition of the operator with given $(z_0,k_0,v_0)$. Hence, the above inequality becomes
	\begin{eqnarray*}
		B(z_0,k_0,v_0)<(\sfT B)(z_0,k_0,v_0)+\epsilon.
	\end{eqnarray*}
	Letting $\epsilon\to0$ in the above yields $B(z_0,k_0,v_0)\le (\sfT B)(z_0,k_0,v_0)$.
	
	Now we show $(\sfT B)(z_0,k_0,v_0)
	\le  B(z_0,k_0,v_0)$.  For any $\epsilon>0$,
	there exists a feasible policy $(b^{\prime }_0,d_0,i_0)$ together with $w(x'_0)$
	to the operator problem
	(\ref{eq:dual-operator}) such that
	\begin{eqnarray}
	&& (\sfT B)(z_0,k_0,v_0) \
	<  \ (1-\tau)F(z_0,k_0)+\tau\delta k_0-d_0-i_0+\tau b'_0 \notag\\
	&&\qquad\quad+\frac{1-\tau}{1+\rho}
\Big\{b'_0 1_{\{b'_0\le0\}}+\big[b^{\prime}_0\mathbf{\mathsf{E}}_{z_0}1_{\{w(x^{\prime }_0)> 0\}}
+L(k^{\prime }_0)\mathbf{\mathsf{E}}_{z_0}1_{\{w(x^{\prime }_0)=0\}}\big]1_{\{b'_0>0\}}\Big\}
+\epsilon,\label{eq-lem42-proof1}\\
	&&b'_0\le B(z'_0,k'_0,w(x'_0)).\notag
	\end{eqnarray}
	The last inequality implies
	\begin{eqnarray*}
		V(z'_0,b'_0,k'_0)\ge  V(z'_0,B(z'_0,k'_0,w(x'_0)),k'_0)
		=w(x'_0),
	\end{eqnarray*}
	where the first inequality is due to $V(z'_0,b,k'_0)$ being decreasing in $b$
	and the second equality is due to $V(z,b,k)$ being the inverse function of
	$B(z,k,\cdot)$.
	In view of the condition (\ref{eq-upperbound-q}) and the above inequality,
	we can rewrite the inequality (\ref{eq-lem42-proof1}) as
	\begin{eqnarray*}
		&& (\sfT B)(z_0,k_0,v_0) \
		<  \ (1-\tau)F(z_0,k_0)+\tau\delta k_0-d_0-i_0+\tau b'_0\notag\\
		&&\qquad\quad+\frac{1-\tau}{1+\rho}
\Big\{b'_0 1_{\{b'_0\le0\}}+\big[b^{\prime}_0\mathbf{\mathsf{E}}_{z_0}1_{\{V(x^{\prime }_0)> 0\}}
+L(k^{\prime }_0)\mathbf{\mathsf{E}}_{z_0}1_{\{V(x^{\prime }_0)=0\}}\big]1_{\{b'_0>0\}}\Big\}
+\epsilon.\\
	\end{eqnarray*}
	Thus, we have established $(b'_0,d_0,i_0)$ is a feasible policy for
	the dual problem (\ref{eq:FirmBellman-s-dual}). Then, the above inequality implies
	\begin{eqnarray*}
		&& (\sfT B)(z_0,k_0,v_0)
		<  B(z_0,k_0,v_0)+\epsilon.
	\end{eqnarray*}
	Letting $\epsilon\to0$ in the above concludes the proof.
\endproof

\proof{Proof of Lemma~\ref{lem-r-bound}.}
\label{proof:lem-r-bound}
	We start with establishing some useful inequalities.
	In view of the bound (\ref{eq-upperbound-q}) given by the break-even condition,
	we have for any $y\in\cY$ and $a\in \cA(y)$,
	\begin{eqnarray}
	r(y,a)&\le&(1-\tau)F(z,k)+\tau\delta k-d-i+\left(\tau
+\frac{1-\tau}{1+\rho}\right)b'\label{eq-boundfor-r-0}\\
	&\le&(1-\tau)F(z,k)+\tau\delta k+ \eta_1 \notag\\
	&\le&\left[(1-\tau)\bar{z} A +\tau\delta\right] k+ \eta_1+
	(1-\tau)\bar{z}A,
	\label{eq-boundfor-r}
	\end{eqnarray}
	where
	\begin{eqnarray*}
		\eta_1=\frac{1+\rho\tau}{1+\rho}\bar{b},
	\end{eqnarray*}
and we used $k^\alpha\le 1+k$ for  all $k\ge0$ in obtaining the last inequality.
	In view of the first constraint for the operator $\sfT$ definition
	(which is equivalent to $r(y,a)\ge0$)
	and the inequality (\ref{eq-boundfor-r-0}), we have
	\begin{eqnarray}
	i&\le&(1-\tau)\bar{z} A k^\alpha +\tau\delta k+ \eta_1.\label{eq-boundfor-i}
	\end{eqnarray}
	
Now, let
	\begin{eqnarray*}
		\phi(z,k,v)\equiv\phi_0(k)= \left[(1-\tau)\bar{z} A +\tau\delta\right] k+\eta.
	\end{eqnarray*}
	For any given $\epsilon>0$, we will choose $\eta$ such that the inequalities
	in the lemma holds.
	
	The first inequality in the lemma clearly follows from the inequality (\ref{eq-boundfor-r}) if we choose
	$\eta\ge \eta_1+(1-\tau)\bar{z}A$.
	
	Consider the second inequality in the lemma. In view of $k'=(1-\delta)k+i$,
	it follows from the inequality (\ref{eq-boundfor-i}),
	\begin{eqnarray*}
		\phi_0(k')&=&\left[(1-\tau)\bar{z} A +\tau\delta\right] k'+\eta\notag\\
		&\le&\left[(1-\tau)\bar{z} A +\tau\delta\right]
		\big\{ (1-\tau)\bar{z}A k^\alpha + [1-\delta(1-\tau)]k+\eta_1\big\}\notag
		+\eta.
	\end{eqnarray*}
	With the above inequality, in order for $\phi(x')\le (1+\epsilon) \phi(x)$, it suffices
	to have
	\begin{eqnarray*}
		\left[(1-\tau)\bar{z} A +\tau\delta\right]
		\big\{[\epsilon+\delta(1-\tau)]k -(1-\tau)\bar{z}A k^\alpha-\eta_1\big\}
		+\epsilon\eta\ge0.
	\end{eqnarray*}
	It is elementary to show that
	\begin{eqnarray*}
		m(\epsilon)=\min_{k\ge0}
		\big\{[\epsilon+\delta(1-\tau)]k -(1-\tau)\bar{z}A k^\alpha-\eta_1\big\},
	\end{eqnarray*}
	is a finite negative number. Hence, the above inequality holds if
	$\eta \ge - m(\epsilon)\left[(1-\tau)\bar{z} A +\tau\delta\right]/\epsilon$.
	The lemma is proved by choosing
	\begin{eqnarray*}
		\eta=\max\left\{ \eta_1+(1-\tau)\bar{z}A,-\frac{m(\epsilon)\left[(1-\tau)\bar{z} A +\tau\delta\right]}{\epsilon}\right\}.
	\end{eqnarray*}
\endproof

\proof{Proof of Lemma~\ref{lem-T-discounting}.}
\label{proof:lem-T-discounting}
	We note that
	\begin{eqnarray*}
		&& \sfT (f+a\phi)(z,k,v)=\sup_{(b^{\prime },d,i)}\Big\{
		(1-\tau)F(z,k)+\tau\delta k-d-i+\tau b'+\notag\\
		&&\phantom{\sfT (f+a\phi)(z,k,v)=\sup_{(b^{\prime },d,i)}\Big\{}
		+\frac{1-\tau}{1+\rho}\big[b' 1_{\{b'\le 0\}}
+ h(z,b',k')b' 1_{\{b'>0\}}\big] \Big\},\\
		&&\mbox{subject to}  \notag \\
		&&\qquad (1-\tau)F(z,k)+\tau\delta k-d-i+\tau b'
 +\frac{1-\tau}{1+\rho}\big[b' 1_{\{b'\le 0\}}
+ h(z,b',k')b' 1_{\{b'>0\}}\big]
\ge0,\\
		&&\qquad v=d+\beta \mathbf{\mathsf{E}}_z  w(x'),\notag \\
		&&\qquad b'\le f(z',k',w(x'))+a\phi(z',k',w(x'))
		\mbox{ for all }x'\in\cX\mbox{ such that }w(x')\ge0\mbox{ for all }x'\in\cX,\notag\\
		&&\qquad k'=(1-\delta)k+i,\notag
		\\
		&&\qquad \underbar{$b$}\le b^{\prime }\leq \bar{b},d\geq 0,i\geq 0,
	\end{eqnarray*}%
	where
	\begin{eqnarray*}
		h(z,b',k') =
	\frac{b^{\prime}\mathbf{\mathsf{E}}%
			_{z}1_{\{w(x^{\prime })> 0\}}+L(k^{\prime })\mathbf{\mathsf{E}}%
			_{z}1_{\{w(x^{\prime })=0\}}}{b'}.
	\end{eqnarray*}
	We note that $h(z,b',k')$ is decreasing in $b'$ and $h(z,b',k')\le
	1$ for $b'\ge0$ due to the break-even condition
	(\ref{eq-upperbound-q}).
	
	Next, we note $b' 1_{\{b'\le0\}}=\min\{0,b'\}$ and $b' 1_{\{b'>0\}}=\max\{0,b'\}$.
	The distinction between $ \sfT (f+a\phi)$ and  $\sfT f$ is in the third constraint,
	the former having $b'\le f(z',k',w(x'))+a\phi(z',k',w(x'))$ or $b'-a\phi(z',k',w(x'))
	\le f(z',k',w(x'))$, and the latter having $b'\le f(z',k',w(x'))$.
	
	The last two terms in the above objective function can be written as (with $\phi'$ short for $\phi(z',k',w(x'))$),
	\begin{eqnarray}
	&&\tau b'+\frac{1-\tau}{1+\rho}[
	\min\{0,b'\}+h(z,b',k')\max\{0,b'\}]\notag\\
	&=&\tau (b'- a\phi')+\frac{1-\tau}{1+\rho}[
	 \min\{0,b'-a\phi'\}+ h(z,b'-a\phi',k')\max\{0,b'-a\phi'\}]\notag\\
	&&+\tau a\phi'+\frac{1-\tau}{1+\rho}\Big\{
	\big[ \min\{0,b'\}-\min\{0,b'-a\phi'\}\big]\notag\\
	&&+\big[h(z,b',k')\max\{0,b'\}-h(z,b'-a\phi',k')\max\{0,b'-a\phi'\}\big]\Big\}\notag\\
	&\le&\tau (b'- a\phi')+\frac{1-\tau}{1+\rho}
\big(
 \min\{0,b'-a\phi'\}+ h(z,b'-a\phi',k')\max\{0,b'-a\phi'\}\big)\notag\\
	&&+\tau a\phi'+\frac{1-\tau}{1+\rho} a\phi',
	\label{eq-mono-proof1}
	\end{eqnarray}
	where the last inequality follows from considering each of the three cases $b'\le0$,
	$0\le b'\le a\phi'$ and $b'\ge a\phi'$ to establish
	\begin{eqnarray*}
		&&\Big\{\big[ \min\{0,b'\}-\min\{0,b'-a\phi'\}\big]\notag\\
		&&\qquad+\big[h(z,b',k')\max\{0,b'\}-h(z,b'-a\phi',k')\max\{0,b'-a\phi'\}\big]\Big\}
\le a\phi'.
	\end{eqnarray*}
	Let
	\begin{eqnarray*}
		\theta = (1+\epsilon)\left[\tau +\frac{1-\tau}{1+\rho}\right];
	\end{eqnarray*}
	we choose $\epsilon>0$ small enough to ensure $\theta<1$. Then,
	combining the inequality (\ref{eq-mono-proof1})
	and the second inequality in Lemma~\ref{lem-r-bound} and , we have
	\begin{eqnarray*}
		&&\tau b'+\frac{1-\tau}{1+\rho}[
	\min\{0,b'\}+ h(z,b',k')\max\{0,b'\}]\notag\\
		&\le&\tau (b'- a\phi')+\frac{1-\tau}{1+\rho}
[ \min\{0,b'-a\phi'\}+ h(z,b'-a\phi',k')\max\{0,b'-a\phi'\}]
+\theta a\phi.
	\end{eqnarray*}
	In view of the definition for the operator, the above inequality establishes the lemma.
\endproof

\section{Proofs of propositions in section~\ref{sec-capital-debt-dividend}}
\label{proof:sec-capital-debt-dividend}

\proof{Proof of Proposition~\ref{pro-monotone}.}
\label{proof:pro-monotone}	
	\textbf{(a)} This follows immediately from Theorem~\ref{thm-main-theorem} and Lemma~\ref{lem-monotone}.

\noindent
	\textbf{(b)} The monotonicity result
	clearly holds for $b'\le0$. For $b'>0$, we note that the comment following
	the break-even condition (\ref{eq:BE_basic-s}) that $q(z,b',k')$ is
	increasing in the value function. Then the monotonicity result follow from
	part (a) (with respect to $b'$).

\noindent
\textbf{(c)}
		When $b'\le 0$, we have $q(z,b',k')=1/(1+\rho)\equiv q$.
If there exists an optimal policy with $b'<0$ for a given state
	$(z,b,k)\in\cX_{feas}$, then for any
	$\epsilon>0$, there exists a feasible policy $(b',d_\epsilon,i_\epsilon)$
	such that
	\begin{equation}
		V(z,b,k)<d_\epsilon+\beta \ex_z
		V(z',b',k'_\epsilon)+\epsilon,\label{eq-V-bound-in-s2}
	\end{equation}
	where $k'_\epsilon=(1-\delta)k+i_\epsilon$.
	It is immediate	to verify that
	$(b'-\Delta,d'_\epsilon+[\tau+(1-\tau)q]\Delta,i_\epsilon)$ for any
	$\Delta>0$ is
	a feasible policy for the state $(z,b,k)$; hence,
	\begin{eqnarray}
		V(z,b,k) &\ge& d'_\epsilon+[\tau+(1-\tau)q]\Delta
		+\beta\ex_z V(z', b'-\Delta,k'_\epsilon),\label{eq-V-bound-in-s3}
	\end{eqnarray}
where the inequality follows from the Bellman equation (\ref{eq:FirmBellman-s}).

Now consider $(z',b',k'_\epsilon)$. It follows from the Bellman equation (\ref{eq:FirmBellman-s})
that there exists a feasible policy
	$(b''_\epsilon,d'_\epsilon,i'_\epsilon)$ with $d'_\epsilon>0$	 such that
	\begin{equation*}
		V(z',b',k'_\epsilon)<d'_\epsilon+\beta \ex_{z'} V(z'',b''_\epsilon,k''_\epsilon)+\epsilon,
	\end{equation*}
	where $k''_\epsilon=(1-\delta)k'_\epsilon+i'_\epsilon$. It is immediate to check that
	$(b''_\epsilon,d'_\epsilon-\Delta,i'_\epsilon)$ is a feasible solution for the state
	$(z',b'-\Delta,k'_\epsilon)$ when $0<\Delta<d'_\epsilon$; hence,
	\begin{equation*}
		V(z',b'-\Delta,k'_\epsilon)\ge d'_\epsilon-\Delta+ \beta \ex_{z'}
		V(z'',b''_\epsilon,k''_\epsilon).
	\end{equation*}
	Combining the above two inequalities yields,
	\begin{equation*}
		V(z',b',k'_\epsilon)<V(z',b'-\Delta,k'_\epsilon)+\Delta+\epsilon;
	\end{equation*}
this, together with the inequality (\ref{eq-V-bound-in-s3}), yields,
\begin{eqnarray*}
		V(z,b,k) &>&d'_\epsilon+[\tau+(1-\tau)q]\Delta
		+\beta\left[\ex_z V(z', b',k'_\epsilon)-\Delta-\epsilon\right].
	\end{eqnarray*}
The latter inequality together with the inequality (\ref{eq-V-bound-in-s2}) gives
\begin{eqnarray*}
		V(z,b,k)
		&>&[\tau+(1-\tau)q-\beta]\Delta + V(z,s,k)-(1+\beta)\epsilon,
	\end{eqnarray*}
	The above inequality implies that
	$[\tau+(1-\tau)q-\beta]\Delta<(1+\beta)\epsilon$ for all
	$\epsilon>0$,	clearly contradicting  $\tau+(1-\tau)q\ge\beta$ by noting that $q=\beta$.
\endproof

\proof{Proof of Proposition~\ref{pro-moderate-capital}.}
The existence of a finite optimal solution (\ref{eq-optimaltarget}) is because the first two terms
in the maximand are concave and have a a finite optimal solution and the third in the maximand is
bounded.

The optimal target case corresponds
the case when the Bellman equation (\ref{eq:FirmBellman-s}) is achieved
with the constraints $d\ge0$ and $i\ge0$ not strictly binding. In this case, we can write
\begin{eqnarray}
V(x)&=&R(z,k)-b+h(z),\label{eq-fullmodel-v1}
\end{eqnarray}
where
\begin{eqnarray*}
	h(z)&=& \max_{0\le b'\le\bar{b},k'}
	\big\{ [\tau+(1-\tau)q(z,b',k')]b'-k'+\beta\ex_z V(x')\big\}.
\end{eqnarray*}
(Given the existence of the equilibrium, the $\sup$ is clearly achieved, so we will use $\max$.)
Substituting the value function (\ref{eq-fullmodel-v1}) into
the Bellman equation (\ref{eq:FirmBellman-s}) yields,
\begin{eqnarray}
h(z)&=&\beta\ex_z h(z')\notag\\
&&+\max_{k'}\Big\{ \beta(1-\tau)\ex_z F(z',k')
-\big[1-\beta+\beta\delta(1-\tau)\big] k'+N(z,k')\Big\}.\label{eq-fullmodel-v2}
\end{eqnarray}
The optimal solution $k'$ in the equality (\ref{eq-fullmodel-v2}) is given by
$k'=k^*_d(z)$. Under the assumption of the proposition, clearly, $i=k^*_d(z)-(1-\delta)k$ is the optimal investment for the firm, and the remainder of the proof is straightforward.
\endproof

\section{Proof of proposition in section~\ref{sec:extend}}
\label{proof:sec:extend}

This subsection is mainly on
the proof of Proposition~\ref{thm-main-proposition} (Exogenous Cash Flow Model).
The proof is similar to the proof of Theorem~\ref{thm-main-theorem},
so we outline without getting into the details.

First, the prime problem is given by (c.f.\ (\ref{prime-operator1})),
\begin{eqnarray}
\label{eq:FirmBellman-s-prime-exogen}
V(x)&=&\sup_{\underbar{${\scriptstyle b}$}\le b^{\prime }\leq \bar{b},d\geq 0}
\quad \big\{d
+\beta
\mathbf{\mathsf{E}}_z V(s')\big\}, \\
&&\mbox{subject to} \quad d= z-b +\tau b'+\frac{1-\tau}{1+\rho}
\Big\{b' 1_{\{b'\le0\}}+ b^{\prime}1_{\{b'>0\}}\mathbf{\mathsf{E}}_{z}1_{\{V(x^{\prime })> 0\}}
\Big\}\notag
\end{eqnarray}%
for $x\in \mathbb{X}_{feas}$, and $V(x)=0$ for  $x\not \in \mathbb{X}$.

Now we consider a two-dimensional dual space $\cY=\cZ\times \Re_+$.
For $y=(z,v)\in\cY$; its first component $z$ denotes the exogenous state, and
its second component $v$ denotes the value of the firm.
Given an equilibrium condition, we define the dual problem for (\ref{eq:FirmBellman-s-prime-exogen})
as follows:
\begin{eqnarray}
\hat{B}(z,v)&=&\sup_{\underbar{${\scriptstyle b}$}\le b^{\prime }\leq \bar{b},d\geq 0}
\left\{
z-d+\tau b'
+\frac{1-\tau}{1+\rho}
\Big\{b' 1_{\{b'\le0\}}+b^{\prime}1_{\{b'>0\}}
\mathbf{\mathsf{E}}_{z}1_{\{V(x^{\prime })> 0\}}\Big\}
\right\} \label{eq:FirmBellman-s-dual-exogen}\\
\mbox{subject to}&& z-d+\tau b'
+\frac{1-\tau}{1+\rho}
\Big\{b' 1_{\{b'\le0\}}+b^{\prime}1_{\{b'>0\}}
\mathbf{\mathsf{E}}_{z}1_{\{V(x^{\prime })> 0\}}\Big\}\ge0,\notag\\
&&v=d+\beta \mathbf{\mathsf{E}}_z  V(x'),\notag
\end{eqnarray}%
where we note $x'=(z',b')$.

Following almost the same line of the proof, we can show that a version of part (a) of
Proposition~\ref{pro-monotone} holds in this case; that is, $V(z,b)$ is strictly increasing in
$b\in (\underbar{$b$}(z),\bar{b}(z))$ for all $x=(z,b)\in \mathbb{X}_{feas}$.
Let $B(z,v)$ be the inverse of the equilibrium value function $V$ in $b$;
i.e., $V(z,B(z,v))\equiv v$ for all $y\in \mathbb{Y}$ and $B(z,V(z,b))\equiv b$
for all $b\in (\underbar{$b$}(z),\bar{b}(z))$. We have a version of Lemma~\ref{lem-dual-equiv}.

\begin{lemma}
	\label{lem-dual-equiv-exogen}
	$B(z,v)=\hat{B}(z,v)$.
\end{lemma}

Next, we similarly define an equilibrium operator $\sfT$
on all continuous function $f$ of $y=(z,v)\in\cY$ as follows,
\begin{eqnarray}
(\sfT f)(z,v)&=&\sup_{\underbar{${\scriptstyle b}$}\le b^{\prime }\leq \bar{b},d\geq 0}
\left\{
z-d+\tau b'
+\frac{1-\tau}{1+\rho}
\Big\{b' 1_{\{b'\le0\}}+ b^{\prime}1_{\{b'>0\}}
\mathbf{\mathsf{E}}_{z}1_{\{w(x^{\prime })> 0\}}
\Big\}\right\}
\label{eq:dual-operator-exogen}\\
\mbox{subject to}&& z-d+\tau b'
+\frac{1-\tau}{1+\rho}
\Big\{b' 1_{\{b'\le0\}}+ b^{\prime}1_{\{b'>0\}}
\mathbf{\mathsf{E}}_{z}1_{\{w(x^{\prime })> 0\}}\Big\}\ge0,
 \notag\\
&&\qquad v=d+\beta \mathbf{\mathsf{E}}_z  w(x'),\notag \\
&&\qquad b'\le f(z',w(x'))\mbox{ for all }x'\in\cX\mbox{ such that }w(x')\ge0
\mbox{ for all }x'\in\cX.\notag
\end{eqnarray}%

A version of Lemma~\ref{lem-BisFixedPT} holds here as well (whose proof is also similar).
\begin{lemma}
	\label{lem-BisFixedPT-exogen}
	Any equilibrium $B(z,k,v)$ is a fixed point of $\sfT$.
\end{lemma}

What remains is to show that the operator  $\sfT$ has a unique fixed point.
The key component for this proof is a version of Lemma~\ref{lem-r-bound}, which gives
an upper bound for the objective function in (\ref{eq:dual-operator-exogen}).

 Let $\cA(z,v)$ denote the set of the feasible policy for the operator problem
(\ref{eq:dual-operator-exogen}); that is, the set of all $(b',d)$ that satisfies
the constraints of the optimization problem (\ref{eq:dual-operator-exogen}).
For any $y=(z,v)\in\cY$ and $a=(b',d)\in \cA(z,v)$,
we the objective function in (\ref{eq:dual-operator-exogen})
(which we also call the (single-period) reward function) by
\begin{eqnarray*}
	r(y,a)&=&
z-d+\tau b'
+\frac{1-\tau}{1+\rho}
\Big\{b' 1_{\{b'\le0\}}+ b^{\prime}1_{\{b'>0\}}
\mathbf{\mathsf{E}}_{z}1_{\{w(x^{\prime })> 0\}}
\Big\},
\end{eqnarray*}
where the state $x'=(z',b')$. The version of Lemma~\ref{lem-r-bound} follows (whose proof is obvious).

\begin{lemma}
	\label{lem-r-bound-exogen}
	For any $\epsilon>0$, there exists an  $\eta>0$ such that the function,
	\begin{eqnarray*}
\phi(y)\equiv		
\phi(z,v)\equiv \phi_0 =\bar{z}+\eta,
	\end{eqnarray*}
	ensures that for all $y\in\cY$,
	\begin{eqnarray*}
		r(y,a)\le \phi(y)\mbox{ for all }a\in \cA(y)\qquad\mbox{and}
		\qquad \phi(y') \le (1+\epsilon) \phi(y) \mbox{ for all } a\in \cA(y).
	\end{eqnarray*}
\end{lemma}

The remaining proof follows the same line as in the proof of Theorem~\ref{thm-main-theorem}.

\begin{singlespace}
	\bibliographystyle{aer}
	\bibliography{Biblio_Defaultable_2022_Feb}
\end{singlespace}

\end{document}